# Compact and robust design of the optical system for cold atom interferometer in space

Danfang Zhang[1,2], Jinting Li[1,2], Wenzhang Wang[1,2], Weihao Xu[1,2], Jie Fang[1], Xiao Li[1], Qunfeng Chen[1], Yibo Wang[1], Biao Tang[1], Lin Zhou[1,3], Jiaqi Zhong[1,3,5], Xi Chen[1,4,6], Jin Wang[1,3,4,7] and Mingsheng Zhan[1,3,4]

[1]*State Key Laboratory of Magnetic Resonance and Atomic and Molecular Physics, Innovation Academy for Precision Measurement Science and Technology, Chinese Academy of Sciences, Wuhan 430071, China*

[2]*School of Physical Sciences, University of Chinese Academy of Sciences, Beijing 100049, China*

[3]*Hefei National Laboratory, Hefei 230088, China*

[4]*Wuhan Institute of Quantum Technology, Wuhan 430206, China*

[5]*jqzhong@apm.ac.cn*

[6]*chenxi@apm.ac.cn*

[7]*wangjin@apm.ac.cn*

**Abstract**：    The optical system is a complex and precise subsystem for the atom interferometer (AI), especially for those used in field or space applications. Here, we introduce the design of the optical system of the China Space Station atom interferometer (CSSAI). The scheme is optimized to reduce the complexity while maintaining the capability to achieve the dual-species AI. It features a fused silica optical bench with bonding technology, ensuring compactness and smaller thermal deformation. Spatial structures are designed to isolate the vibration and transfer the heat. After assembling, the optical system has a size of 250 mm × 240 mm × 104 mm and weighs 5.2 kg. After installing in the CSSAI, it passed the thermal and mechanical tests and then launched to the China Space Station (CSS). The output laser power changes are less than 15% from ground to space, and its long-term fluctuations are less than 2.5% for months in space. Cold atom preparation and interference are also realized in space. This optical system is extremely integrated and robust, which provides a foundation for the design of future cold atom payloads in space.

## 1. Introduction

The atom interferometer (AI) is a high-precision inertial measurement instrument [1]. AI-based gravimeters [2-8], gradiometers [9-11] and gyroscopes [12-15] have been developed and AI has wide applications in fundamental physics research [16-25]. The measurement resolution of AI is proportional to the square of its interference time. On the ground, the prepared cold atom cloud will free-fall under the Earth's gravity. While in space, it will be at rest in the vacuum chamber for a long time. Consequently, the interference time will be much longer and the measurement precision will be much higher for AI in space. Several AI projects have been proposed for space applications, ranging from gravity mapping [26-28] to fundamental physics research [29]. Preliminary studies have also been carried out in microgravity platforms, such as drop tower [30], sounding rocket [31], 0g plane [32] and space station [33].

The optical system is one of the key sub-systems of AI and provides lasers for cold atom manipulation. This system requires extremely high structural stability due to its precise optical components. How to make the optical system compact and stable is a hot topic for applications of AI [34-48]. In space, the optical system faces additional challenges, including limited resources, space environment adaptation, and long-term work reliably. Several optical systems designed for cold atom experiments in space have been demonstrated. The space cold atom clock project ACE / PHARAO designed an optical system with compact design and precision temperature control [49]. The cold atom clock missions in the China space station (CSS) designed a compact optical system using aluminum silicon carbide as the optical bench [50, 51]. The Cold Atom Laboratory (CAL) in the International Space Station adopted an optical design with commercially available fiber optical components [52]. The sounding rocket project MAIUS used the Zerodur as the optical bench and developed integrated optical components [39, 53]. The Cold Atom Physics Research Rack (CAPR) in the China space station adopted an all-fiber optical system design, which includes 780 nm lasers for rubidium atom preparation and also high-power 1064 nm lasers for the study of ultracold degenerate gases [54, 55].

The China Space Station Atom Interferometer (CSSAI) is an integrated AI installed in the CSS in 2022. It aims to realize $^{87}$Rb and $^{85}$Rb dual-species AI in space and carry out experiments such as gravity measurement and equivalence principle (EP) test [56]. To improve its microgravity level, CSSAI is installed inside the Free-floating Platform for Microgravity Experiment (FPME) of the High Microgravity Level Research Rack (HMLR) [57, 58], as shown in Fig. 1. Because of the limited resources that HMLR can provide, there are strict constraints on CSSAI's volume, weight, and power consumption. Additionally, the lack of an effective heat dissipation channel presents a critical challenge, as it leads to a continuous rise in the CSSAI's temperature, preventing it from achieving thermal equilibrium during operation.

These constraints challenge the payload's design, especially its complex and sophisticated optical system. Firstly, the optical system must generate the required lasers for dual-species AI with a small volume, lightweight design, and low power consumption. Secondly, its thermal deformation must be small under unstable ambient temperatures. Thirdly, its mechanical deformation must be small due to gravity changes from Earth to space, and it must be robust enough to withstand vibrations during the rocket launch. Finally, the optical system has to be reliable during its lifetime under unmanned working conditions.

This article introduces the design and tests of the optical system of CSSAI. A scheme of sideband stabilization, sideband modulation, and tapered amplifier (TA) synchronized amplification is proposed to significantly simplify the optical system. Integrated optical components are designed and low-expansion fused quartz bench with double-sided bonding technology is employed, which enhances compactness, temperature adaptability, and long-term stability. Specific vibration isolation and heat transfer components are designed to solve the problems of fragility and low thermal conductivity of the optical bench, thereby improving its mechanical and thermal environment adaptability. The optical system of CSSAI has a very compact volume of only 6.2 l, which is the most compact one among typical reported optical systems for AIs both on ground and for space applications [34, 38, 39, 42, 44, 46, 48]. It has also passed the mechanical and thermal tests and operated stably for more than 2 years in orbit, which showing its excellent environmental adaptability and reliability.

The article is organized as follows: Section 2 introduces the optical system's design requirements. Section 3 describes its scheme, mechanical design, and bonding technology. Section 4 discusses the simulation of its thermal and gravity-induced deformation. Section 5 presents the optical system's performance and test on the ground and in space. Section 6 gives the conclusion and discussion.

## 2. Requirement of the optical system's design

The FPME provides limited resources for CSSAI's design. The CSSAI must meet size requirements of 460 mm×330 mm×260 mm, and the power consumption should be less than 80 W. The CSSAI comprises the physical, optical, and electronic sub-systems. Therefore, the optical system has even fewer resources available. The temperature-rising rate of CSSAI during operation is about 0.2 °C/min. The optical system must be operated properly under this temperature variation circumstance.

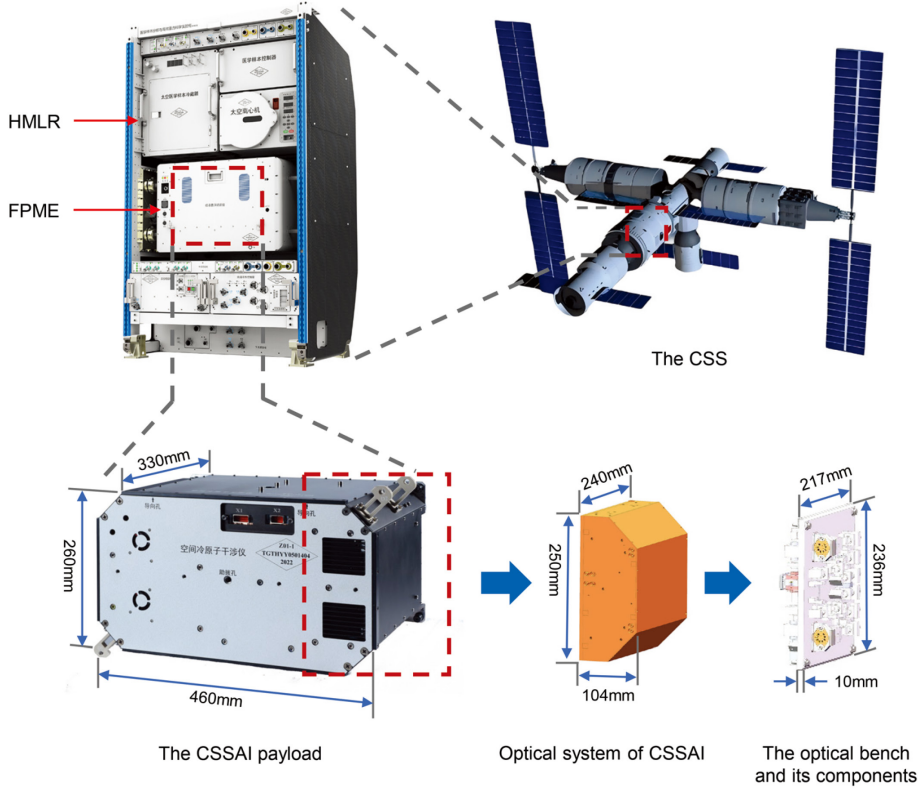

Fig. 1. The installation of the China Space Station atom interferometer (CSSAI) in the High Microgravity Level Research Rack (HMLR), the structure of the CSSAI, and the requirement for the size of its optical system.

Lasers with different frequencies are needed for AI at different stages. This includes laser cooling, state preparation, atom interference, and atom fluorescence detection. For each stage, at least two lasers with different frequencies are needed. We take $^{87}$Rb AI as an example, these two lasers are labeled as $l_{87a}$ and $l_{87b}$ at different stages. Their frequencies are $f_{87a}$ and $f_{87b}$, where $f_{87b} > f_{87a}$. A reference frequency $f_{87ref}$ is defined which is the frequency of the $^{87}$Rb $|5^2S_{1/2}\ F=2\rangle \rightarrow |5^2P_{3/2}\ F'=3\rangle$ transition. The differential frequencies are defined as $\Delta f_{87a}= f_{87a} - f_{87ref}$ and $\Delta f_{87b}=f_{87b} - f_{87a}$, as shown in Fig. 2a. The required values of $\Delta f_{87a}$ and $\Delta f_{87b}$ during the interference time sequence are shown in Fig. 2c. In the cooling stage. $\Delta f_{87a}$ is about -15 MHz and $\Delta f_{87b}$ is 6583 MHz. In the polarization gradient cooling (PGC) process, $\Delta f_{87a}$ is set to -60 MHz and $\Delta f_{87b}$ is set to 6628 MHz. In the state-preparing stage, $\Delta f_{87a}$ is turned to -266 MHz, and $l_{87b}$ is turned off. $l_{87a}$ is in resonance to the $|5^2S_{1/2}\ F=2\rangle \rightarrow |5^2P_{3/2}\ F'=2\rangle$ transition of $^{87}$Rb atoms and pumps atoms from the $|5^2S_{1/2}\ F=2\rangle$ state to the $|5^2S_{1/2}\ F=1\rangle$ state. In the Raman transition stage, $\Delta f_{87a}$ is set to 450 MHz, and $\Delta f_{87b}$ is set to 6834 MHz. $l_{87a}$ together with $l_{87b}$ introduce the Raman transition. In the fluorescence detection stage, $\Delta f_{87a}$ is turned to 0 MHz, and laser $l_{87b}$ is turned off. $l_{87a}$ excites the fluorescence for the atoms in the $|5^2S_{1/2}\ F=2\rangle$ state. A camera collects the fluorescence to obtain the interference fringe [56]. The laser frequency requirement for the $^{85}$Rb AI is similar. To realize the $^{87}$Rb and $^{85}$Rb AI at the same time, the optical system needs at least 4 independent lasers with different frequencies, and their frequencies can be turned for several hundred MHz within tens of milliseconds.

The output laser beams need to have sufficient powers and their amplitudes must be controlled in a precise time sequence for AI. To achieve sufficient cold atom number with finite laser power, the CSSAI uses the 2-dimensional magneto-optical trap (2D-MOT) and 3-dimensional magneto-optical trap (3D-MOT) techniques for laser cooling. Two laser beams are needed for these cooling lasers. To save the laser power and reduce the complexity, the laser beam for the 3D-MOT is also used for atom fluorescence exciting. An additional laser beam is used for the Raman laser. An illustration of the three laser beams for the cold atom manipulation is shown in Fig. 2b. The typical time sequence of the amplitude control of these laser beams for $^{87}$Rb AI is shown in Fig. 2d. The requirement of the amplitude control for the $^{85}$Rb AI is similar.

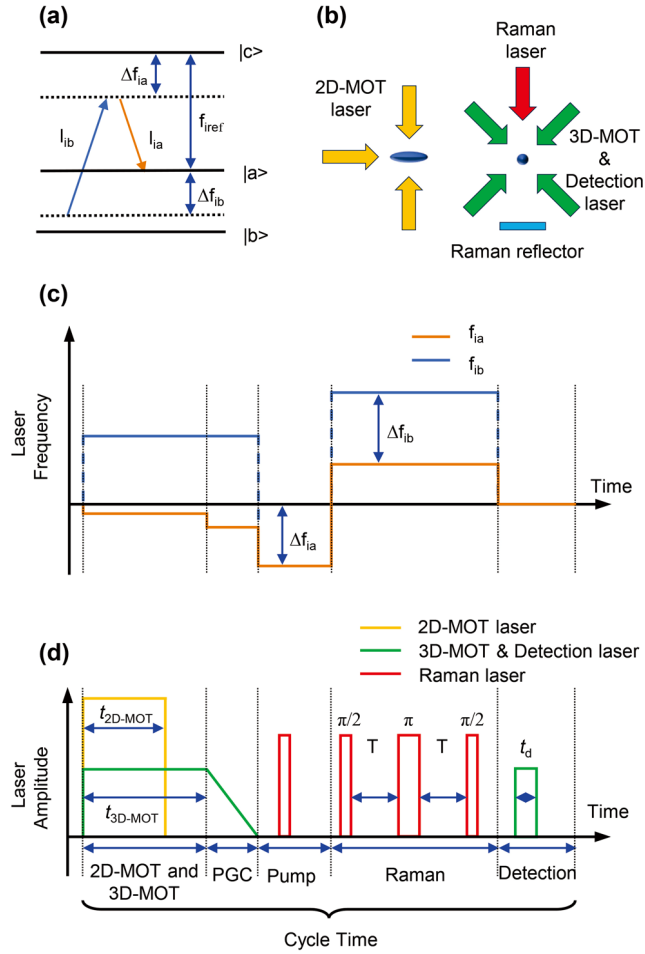

Fig. 2. Laser frequency and amplitude requirements of the optical system of CSSAI. (a) Illustration of the laser transitions for a single species AI. (b) Illustration of the 3 output laser beams that required for atom manipulation in the physics system. (c) and (d) The required time sequences for the frequency and amplitude of the lasers during the interference process for a single species AI.

## 3. Design of the optical system

### 3.1 Scheme design

To realize the above optical functions with the design constraints. The optical system should be designed with a simplified scheme and also a low power consumption. The traditional design schemes for the optical system of AI include the acousto-optic frequency shift (AOFS) scheme [42] and optical beat frequency phase locking (OFPL) scheme [34, 36]. For the AOFS scheme, multiple acousto-optic modulators (AOMs) are needed to generate the various laser frequencies, which make the optical system complex and high-power consumption. The OFPL scheme requires a master laser as the laser frequency reference and at least two slave lasers are needed to generate the required laser frequencies for a single-species AI. So, generally, the optical system requires five laser sources under this scheme for the dual-species AI, which also makes the optical system complex.

The sideband frequency locking (SFL) method [37, 41, 59] locks the laser sideband's frequency to an atom spectrum reference. By varying the modulation frequency of the laser sideband, the carrier laser's frequency can be shifted by GHz within the time scale of milliseconds. Compared to the OFPL method, this reduces the usage of the slave laser. The key optical component of the SFL, which is the fiber electro-optic phase modulator (FEOM), is in low power consumption and easy to integrate. The phase modulation sideband generation (PMSG) method [35, 37, 41, 46] can create laser sidebands conveniently. The key optical component of PMSG is also the FEOM. The frequency and amplitude of the generated laser sidebands can be easily controlled by varying the parameters of the microwave signal applied to the FEOM. Compared to the AOFS method, this method avoids the usage of multiple AOMs and greatly simplifies the optical system. Therefore, our optical system is designed based on these two methods.

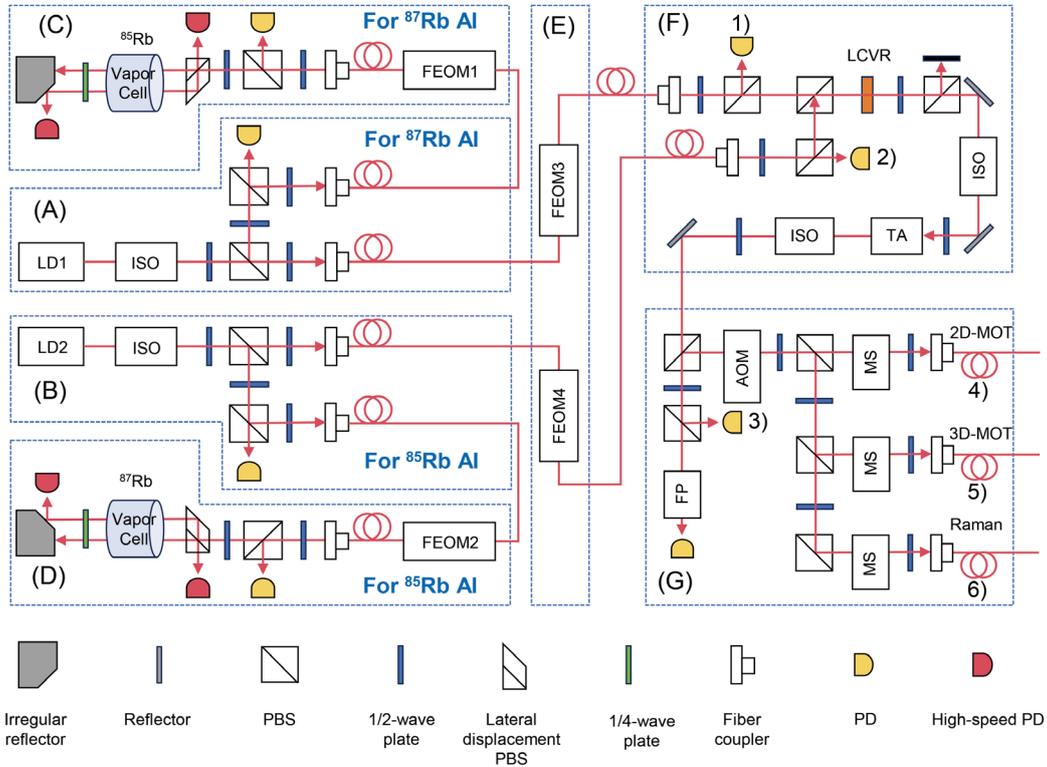

Figure 3. The scheme design of the optical system of CSSAI. (A) and (B) The seed laser generation parts for $^{87}$Rb AI and $^{85}$Rb AI, (C) and (D) The laser frequency stabilization parts for $^{87}$Rb AI and $^{85}$Rb AI, (E) The sideband generation part, (F) The laser power amplification part, (G) The laser power controlling and distribution part. The photo detectors that monitor the laser powers at some key positions is also labeled in this picture, where 1) and 2) are for injection laser powers of TA, 3) is for the output laser power of TA, 4) - 6) are for the output powers for 2D-MOT, 3D-MOT and Raman lasers (The photo detectors are in the physics system and not shown). Abbreviations in the figure: laser diode (LD), isolator (ISO), fiber electro-optic phase modulator (FEOM), liquid crystal variable retarder (LCVR), tapered amplifier (TA), acousto-optic modulator (AOM), mechanical switch (MS), Fabry-Perot cavity (F-P), polarization beam splitter (PBS), photo detector (PD).

The scheme of the optical system is shown in Fig. 3. It includes the seed laser generation, the laser frequency stabilization, the sideband generation, the laser power amplification, and the laser power controlling and distribution parts. For the $^{87}$Rb AI seed laser generation (Fig. 3A), a distributed feedback (DFB) laser diode (EYP-DFB-0780) is used as the seed laser. The output laser is separated into two paths and coupled to two single-mode polarization-maintaining (SMPM) fibers. One fiber is connected to the laser frequency stabilization for $^{87}$Rb AI (Fig. 3C). The laser passes through a FEOM to create sidebands. This modulated laser then passes through an $^{85}$Rb vapor cell to obtain the saturated absorption spectrum. The frequency of the +1 order sideband of the modulated laser is locked to the $|5^2S_{1/2}\ F=3\rangle \rightarrow |5^2P_{3/2}\ F'=co(3,4)\rangle$ transition of the $^{85}$Rb atom. The frequency of the microwave that drives the FEOM ranges from 564 to 1546 MHz. The carrier of the seed laser acts as $l_{87a}$, and the corresponding turning range of $\Delta f_{87a}$ is -400 to +582 MHz. The seed laser generation and the laser frequency stabilization parts for $^{85}$Rb AI have similar configurations (Fig. 3B, 3D). Here, the frequency of the seed laser is $f_{85a}$, and the -1 order sideband of the seed laser is locked to the $|5^2S_{1/2}\ F=2\rangle \rightarrow |5^2P_{3/2}\ F'=co(2,3)\rangle$ transition of the $^{87}$Rb atom. The turning range of $\Delta f_{85a}$ is -284 to +1337 MHz.

The other two fibers from the seed laser generation parts for $^{87}$Rb and $^{85}$Rb AI are connected to the sideband generation part (Fig. 3E). The lasers pass through two FEOMs to create sidebands. The +1 order laser sidebands of the two input lasers act as $l_{87b}$ and $l_{85b}$. The differential frequencies $\Delta f_{87b}$ and $\Delta f_{85b}$ are just the frequencies of the driving microwaves for the FEOMs, which range from 5000 to 7200 MHz and 2700 to 3600 MHz, respectively. The amplitudes of $l_{87b}$ and $l_{85b}$ can be controlled by varying the amplitudes of the respective driving microwave. The output lasers from the sideband generation part are labeled as $l_{87m}$ and $l_{85m}$. The powers of $l_{87m}$ and $l_{85m}$ are about 10 mW and 7 mW, which are not sufficient for cold atom manipulation.

To achieve sufficient laser power while simplifying the optical system and reducing power consumption. These two output lasers are amplified by a single TA. Lasers $l_{87m}$ and $l_{85m}$ are connected to the laser power amplification part (Fig. 3F). They are combined by a polarization beam splitter (PBS) and then pass a liquid crystal variable retarder (LCVR) and another PBS. This retarder is used to adjust the laser power ratio of $l_{87m}$ and $l_{85m}$. The combined laser beam is injected into a TA, and the output power is amplified to about 300 mW. About 1 mW of the output laser power is injected into a Fabry-Perot (F-P) cavity to monitor the power ratio of the amplified lasers of $l_{87m}$ and $l_{85m}$, and also the sideband ratio of each laser. This F-P cavity is carefully designed to distinguish these laser frequencies, as shown in Fig. 4a. The modulation frequencies $\Delta f_{87b}$ and $\Delta f_{85b}$ are 6834 and 3035 MHz during the Raman transition process. The free spectral range of the FP cavity is designed to $f_{FP}=3300$ MHz, which makes $\Delta f_{87b}-2f_{FP}$ and $f_{FP}-\Delta f_{85b}$ approximately the same. For this case, the frequency difference of $l_{87a}$ to $l_{87b}$ and $l_{85a}$ to $l_{85b}$ in the F-P cavity spectrum are similar. As shown in Fig. 4b, all the frequency components of the output lasers are clearly distinguished.

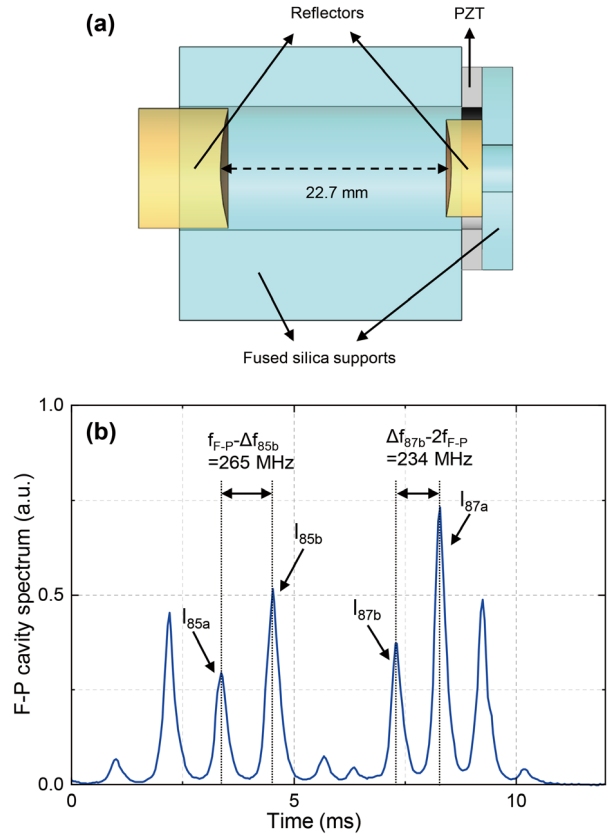

Figure 4. The design of the F-P cavity in optical system and its test result. (a) Structure and material of the F-P cavity. (b) The measured transmission spectrum of the F-P cavity during the Raman transition process when the CSSAI carries out the dual-species AI experiment.

The majority of the output laser of the TA is sent to the laser power controlling and distribution part (Fig. 3G). The laser beam passes through an 80 MHz AOM. The +1 order diffracted laser beam is separated into three paths. Each path passes a mechanical switch (MS) and then is coupled into an SMPM fiber. The total power of the three output lasers is about 120 mW. These fibers are connected to the physics system. After collimation, the three output laser beams are used for the 2D-MOT, 3D-MOT, and Raman transition. By combining the usage of the AOM and MSs, the amplitude of each output laser beam can be controlled with high time precision independently. For example, in the laser cooling stage, the MS for the Raman laser is closed, and in the atom interference stage, the MSs for the 2D-MOT and 3D-MOT lasers are closed. Additionally, the high laser power extinction ratio of the MS minimizes stray light during cold atom manipulation. Besides the five majority parts, several photo detectors are installed at some key positions of the optical system. These detectors monitor laser power and thus the system's health status, which is also shown in Fig. 3.

## 3.2 Mechanical design and assembly

The optical system is characterized by a compact structural design, utilizing fused silica as the optical bench. Compact optical components are designed and bonded onto this bench, offering two significant advantages. Firstly, the fused silica has a very low thermal expansion coefficient, the deformation of the optical system is smaller when it works under unstable environmental temperatures. Secondly, the bonding technique avoids the use of additional structures for position and angle adjusting, which makes the optical system more compact.

The mechanical design of the optical system is shown in Fig. 5a and 5b. The optical bench has a size of 236 mm×217 mm×10 mm, and optical components are bonded on both sides of it. Glass-based components such as the reflection mirror are bonded to the bench directly. Components that cannot be bonded directly, for example, the optical isolator, F-P cavity, AOM, photodetector, and optical fiber coupler, are assembled with specifically designed supports made of fused silica and then bonded to the bench. Power-consuming components, such as the TA laser, DFB laser, and temperature-controlled rubidium vapor cell, are assembled with supports made of invar material and then bonded to the bench. The low thermal expansion coefficient of invar minimizes thermal deformation stress between these supports and the optical bench. All of the optical components are designed with compact size. For example, the size of the reflection mirror is 8 mm×12 mm×20 mm. The height of the laser beams is 15 mm above the optical bench surface on both sides.

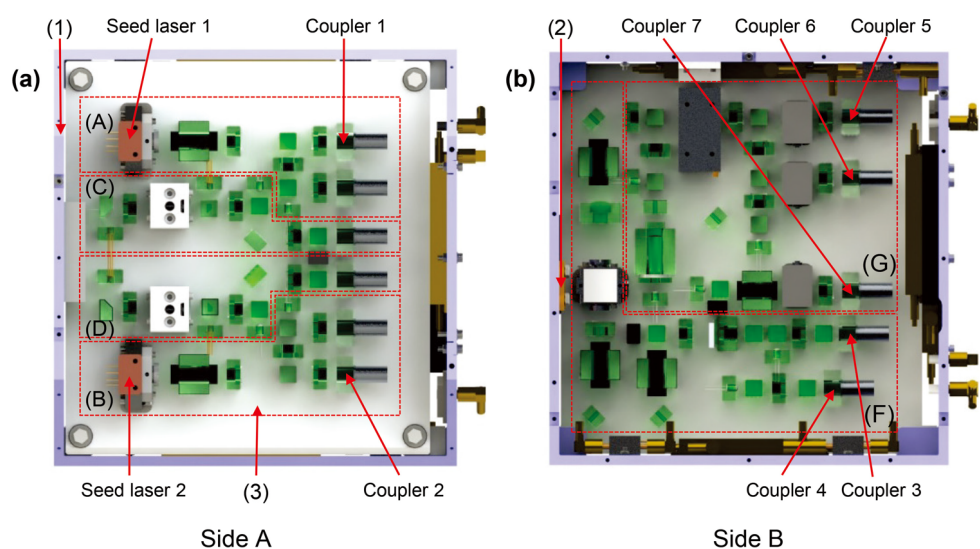

Fig. 5. Mechanical design of the optical system of CSSAI. (a) Side A of the optical system, which contains the seed laser generation parts (A and B) and the laser frequency stabilization parts (C and D). (b) Side B of the optical system, which contains the laser power amplification part (F) and the laser power controlling and distribution part (G). Positions of key optical components for the simulation in section 4 are labeled in this picture, including the two seed lasers and couplers 1-7 which represent the optical fiber couplers for the seed lasers, the optical fiber couplers for the injection lasers of the TA, and the optical fiber couplers of the 3 output lasers of the optical system. Positions of the temperature detectors of the optical system are also labeled in this picture, where (1), (2), and (3) are the positions of the detectors placed at the seed laser1's heat sink, TA's heat sink and the optical bench.

Multiple types of glues are used for bonding. UV light glue (Norland NOA61) is used for bonding between glass components and glass component to the optical bench. The UV light glue has a very low viscosity before the UV curing, which is beneficial for precise adjustment of the optical components [60]. The glue solidifies quickly after UV curing, which shortens the assembly time and improves the bonding efficiency. Two-component epoxy (3M DP2216) is used for bonding between metal and glass materials [61]. The chosen epoxy has a moderate hardness and a high adhesive strength, which guarantees bonding strength while alleviating the thermal expansion mismatch between glass and metal materials. A special assembly device is designed to precisely adjust the position and angle of the optical components. This device has a positioning accuracy of sub-mm for optical components with low assembly precision, such as the mirror, wave plate, and PBS, and has a positioning and angle-adjusting accuracy of μm and μrad for the SMPM fiber coupler, fulfilling the high assembly precision requirement of the optical system.

The optical bench has four holes at its four corners and is fixed to an aluminum alloy frame using four bolts. The rods of the bolts are covered by Teflon tubes for buffering. Silica slices are placed on both sides of the bench in the place of fixed holes for buffering. This protects the fragile glass bench and significantly improves the seismic performance of the optical system. The 4 FEOMs for the SFL and PMSG are mounted on the inner side of the frame to save installation space and conduct their heat. Special heat conducting devices are designed to conduct the heat from the power-consuming optical components on the optical bench to the aluminum alloy frame. These devices have structures where two copper blocks are welded with multiple thin copper braids, which have small thermal resistances and low residual stress. One end of the device is fixed to the power-consuming components, while the other end is fixed to the frame. This effectively transfers the heat from the components to the frame. The TA is the highest power consumption component of the optical system, which has a power of about 2 W. The temperature of the TA is controlled by a TEC, and the heat of the TEC is transferred to the frame through the heat-conducting device. When the ambient temperature is 24°C and the TA is controlled to 28.5°C, the power consumption of TEC is about 1.7 W, which is acceptable for the limited power consumption of CSSAI.

### 4. Thermal and gravity simulation of the optical system

*4.1 The principle of the fiber coupling*

The temperature and gravity changes of the environment will cause deformation of the optical system. The laser is coupled to the SMPM fiber with a mode field diameter of only several micrometers. So, the deformation will lead a significant variation in the fiber coupling efficiency, thereby affecting the output laser power to the physics system. Other laser parameters, such as the laser frequency and sideband ratio, are insensitive to the deformation of the optical system. The principle of the SMPM fiber coupling is shown in Fig. 6. Collimated laser beam is focused by a lens and coupled to the SMPM fiber.

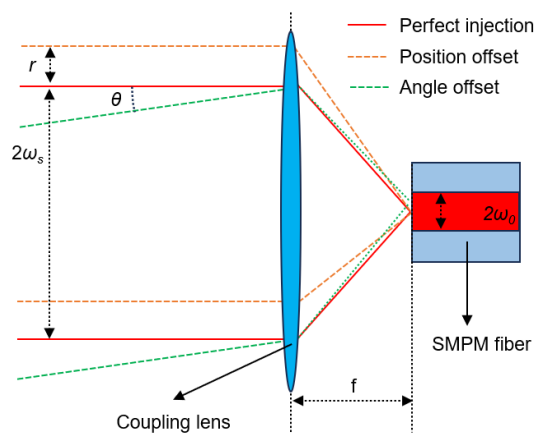

Fig. 6. Schematic of the SMPM fiber coupling.

The coupling coefficient is [62]

$$\eta = \left(\frac{2\omega_a\omega_s}{\omega_a^2+\omega_s^2}\mathrm{Exp}\left[-\frac{r^2}{\omega_a^2+\omega_s^2}\right]\mathrm{Exp}\left[-\frac{\omega_a^2\omega_s^2 k^2 \theta^2}{4(\omega_a^2+\omega_s^2)}\right]\right)^2 \quad (1)$$

where $\omega_a = \lambda f /(\pi\omega_0)$, $\lambda$ is the wavelength of the laser, f is the focal length of the coupling lens, and $\omega_0$ is the fiber's Gaussian-mode-field radius, $\omega_s$ is the waist radius of the incident Gaussian laser beam, $k=2\pi/\lambda$ is the wave vector of the laser. r and θ are the incident laser's position offset and angle offset, as shown in Fig. 6. For our optical system, $\omega_a$ and $\omega_s$ are both about 0.6 mm. Assuming that the r and θ are 10 μm and 100 μrad, the coupling coefficient variations are calculated to be 0.02% and 5.7% respectively, which is mainly determined by the angle offset.

To analyze the power variation quantitatively, finite element analysis is used to simulate the deformation and angle change of the optical system. To make the simulation efficient and accurate, the detailed features of the optical system's components are omitted while maintaining the same dimensions and materials as the real ones. Then the deformations and angle changes of the key optical components, such as the seed lasers, the TA, and the fiber couplers, are evaluated using the simulation.

### 4.2 Thermal Simulation

For the thermal simulation, the initial temperature of the optical system is set the same as the ambient temperature. Then, the temperature controls and all the power-consuming components of the optical system are turned on. For the real experiment, the experiment begins approximately 10 minutes after the power of the optical system is powered on. This time interval allows for the stabilization of the temperature control. Therefore, the simulation calculates the deformation of the optical system after 10 minutes, and uses this as a reference for subsequent deformation analysis.

The simulated temperature distribution of the optical system at 30 minutes is shown in Fig. 7a. The maximum temperature difference of the optical bench is about 4 °C. The relatively large temperature gradient is caused by the low thermal conductivity of fused silica. The component with highest temperature is the AOM, which has a power consumption of about 1 W. The simulated temperatures of several key optical components (labeled in Fig. 5) over time are shown in Fig. 7b, which have temperature rising rates of about 0.2 °C/min.

The simulated thermal induced deformation of the optical system at 30 minutes is shown in Fig. 7c. The maximum deformation of the optical components is about 20 μm, and the maximum deformation of the optical bench is less than 10 μm. The time-dependent angle changes of the key optical components are calculated and shown in Fig. 7d. The maximum angle change of these components is about ±50 μrad after the 20-minute time evolution. Besides these key components, the laser undergoes other components for transmission and reflection, amplifying the angular variation for final fiber coupling. The angle variations of the fiber couplers for the output lasers are expected to exceed 100 μrad after 30 minutes, resulting in a laser power variation of more than 5% according to Eq. (1).

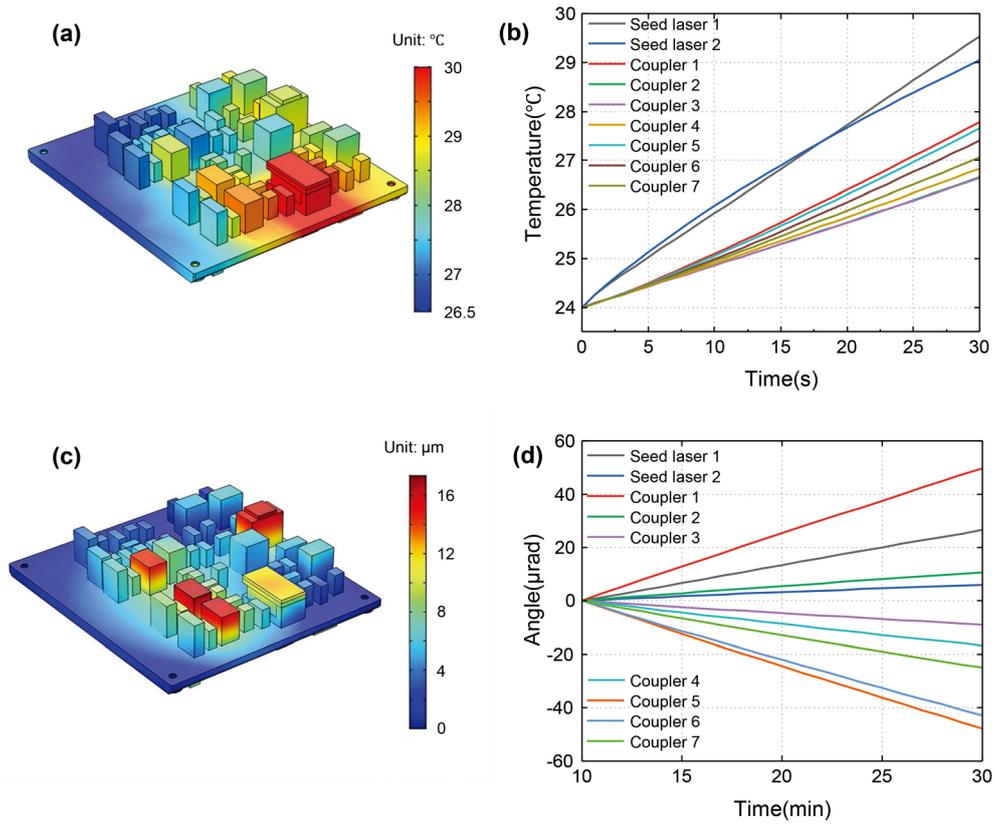

Figure 7. Thermal simulation of the optical system. (a) Simulated temperature distribution of the optical system at the time of 30 minutes. (b) Simulated temperatures of several key optical components over time. (c) Simulated thermal induced deformation of the optical system at the time of 30 minutes. (d) Simulated angle changes of several key optical components over time.

### 4.3 Gravity Simulation

For the gravity-induced deformation simulation, the gravity is in the vertical direction and the optical bench is in the horizontal plane. The deformation of the optical system is zero when the gravity is zero, and the deformation when the gravity changes is calculated. The value of gravity varies from 0 to 10 m/s$^2$, which simulates the transition from microgravity to the Earth's gravity. The simulated deformation of the optical system when the gravity is 10 m/s$^2$ is shown in Fig. 8a. The maximum deformation is less than 1 μm. The angle changes of the key optical components when the gravity changes are shown in Fig. 8b. The maximum angle change of these components is about ±4 μrad when the gravity is 10 m/s$^2$. Similar to the thermal simulation case, beside these key components, other components will amplify the angle variation for the final fiber coupling to the order of about 10 μrad. The power variation of the output lasers is expected to be less than 0.1% during the gravity change, according to Eq. (1).

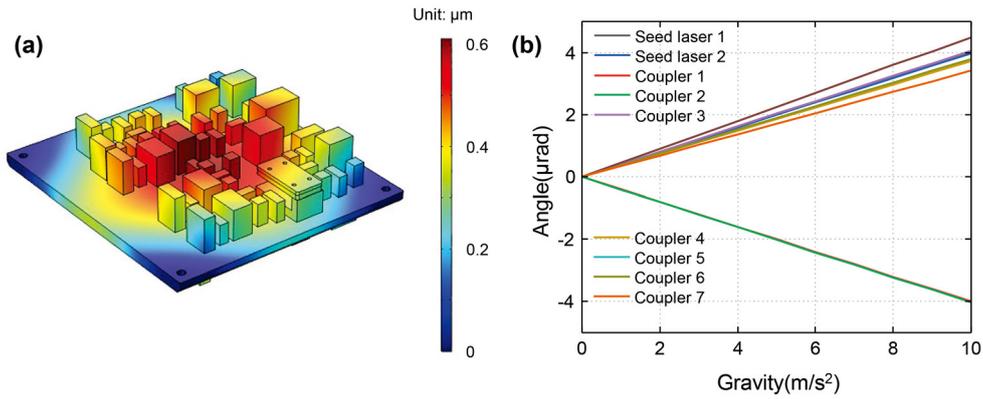

Fig. 8. Gravity-induced deformation simulation of the optical system. (a) Simulated deformation of the optical system when the gravity is 10 m/s2. (b) Simulated angle changes of several key optical components with gravity.

## 5. Performance of the optical system

### 5.1 Ground test

The assembled optical system is shown in Fig. 9. Its size is 250 mm × 240 mm × 110 mm and its mass is 5.2 kg. The power consumption is approximately 15 W during operation. Functions of laser frequency locking and shifting, laser sideband generation, laser power, and laser power ratio adjusting are realized after its assembly. The frequency stabilities of the two frequency-locked seed lasers are better than 1 MHz. The differential frequencies $\Delta f_{87a}$ and $\Delta f_{85a}$ can be adjusted in the range of -0.4 GHz to 0.6 GHz and -0.3 GHz to 1.3 GHz respectively, and the deferential frequencies $\Delta f_{87b}$ and $\Delta f_{85b}$ can be adjusted in the range of 5.0 GHz to 6.9 GHz and 2.7 GHz to 3.6 GHz respectively. The powers of the three output lasers are about 29 mW, 50 mW, and 43 mW for the 3D-MOT, 2D-MOT, and Raman lasers. The laser power ratio of $l_{87}$ to $l_{85}$ can be varied to any value between 0:1 and 10:1. Ground-based experiments of laser cooling and atom interference are also realized by utilizing this compact optical system [56].

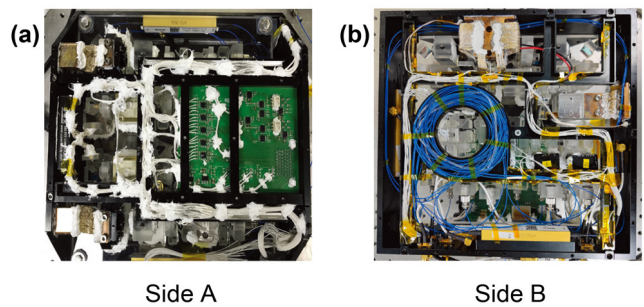

Fig. 9. The assembled optical system (a) Side A, (b) Side B

### 5.2 Environmental test

Environment tests such as thermal and mechanical tests are carried out to test the environmental adaptability of the optical system. For the thermal test, the temperature range is 15 to 32°C. For the mechanical test, the vibration has a standard deviation of 4.28 g from 10 to 2000 Hz. The laser powers at some key positions are recorded before and after these tests. These include the three output lasers of the optical system and the injection and output laser of the TA, which are labeled in Fig. 3. The laser powers are recorded and shown in Table 1. The power variations of the three output lasers are within 10% after the environment tests.

The payload was launched in orbit by the Tianzhou-5 cargo spacecraft in 2022 and then installed in the HMLR of the CSS. The laser powers at the key positions are measured again. The power variations of the three output lasers are within 15% compared to their values before launch. The power variation was larger than the predicted one that is given in Sec. 4.3. This might be caused by a joint effect of vibration during the rocket launch and the gravity change from the ground to space, which cause unexpected deformation of the optical system. However, the power variations are acceptable and do not affect the AI experiment in space.

Table 1. Laser powers recorded at some key positions during the environmental tests and in orbit

| Laser output | Environmental tests on ground | | | | In orbit | |
|---|---|---|---|---|---|---|
| | Before tests (mW) | After the thermal test(mW) | After the vibration test (mW) | Relative laser power variations after the tests | Lasers power (mW) | Relative laser power variations |
| Output laser for 2D-MOT | 50.29 | 44.44 | 49.30 | -1.97% | 43.50 | -11.76% |
| Output laser for 3D-MOT | 29.02 | 26.63 | 26.40 | -9.03% | 26.90 | +1.90% |
| Output laser for Raman transition | 43.04 | 41.96 | 40.46 | -5.99% | 46.18 | 14.09% |
| $^{85}$Rb injection laser of TA | 6.24 | 5.25 | 5.36 | -14.10% | 6.68 | +24.62% |
| $^{87}$Rb injection laser of TA | 9.03 | 7.96 | 7.82 | -13.40% | 9.70 | +24.04% |
| Output laser of TA | 308.69 | 297.16 | 298.72 | -3.23% | 300.20 | +0.50% |

*5.3 Performance in Space*

After the installation of the CSSAI, functional tests and parameter optimization are conducted. Functions of laser frequency locking, laser power and laser ratio controlling, laser sideband generation, and optical switching are realized. Laser cooling and atom interference experiments are carried out, and atom interference for both the $^{85}$Rb and $^{87}$Rb atoms are realized in space [63].

As mentioned in Sec. 2, the temperature of the payload will continue to rise during operation. To protect the payload from overheating, the payload is working intermittently for 50-70 minutes and then naturally cooled to the ambient temperature. For each experiment, it takes approximately 10 minutes to startup and preheat, 20 minutes for scientific experiments, and 20-40 minutes for data transfer. The time sequence is consistent with the thermal simulation in Sec. 4.2. The optical system has three temperature detectors for the heat sinks of the lasers and for the optical bench (labeled in Fig. 5). The temperatures recorded by these detectors during the scientific experiment are shown in Fig. 10a. The temperature rising rates are about 0.2 °C/min, which coincides with the simulation result in Sec. 4.2. The corresponding laser power variations of the output lasers for 2D-MOT and 3D-MOT during the scientific experiment are shown in Fig. 10b. The relative power variations are 24% and 10% respectively. These variations are consistent with simulation results in Sec. 4.2 in terms of magnitude, and are acceptable for the space AI experiment.

The CSSAI has operated normally in orbit for two years. The powers of the output lasers for 2D-MOT and 3D-MOT are recorded for a long time. Fig. 10c shows the powers of these two lasers for 4 months. The standard deviations of the relative powers are 2.5% and 1.4% respectively, which verifies the excellent long-term stability of the optical system.

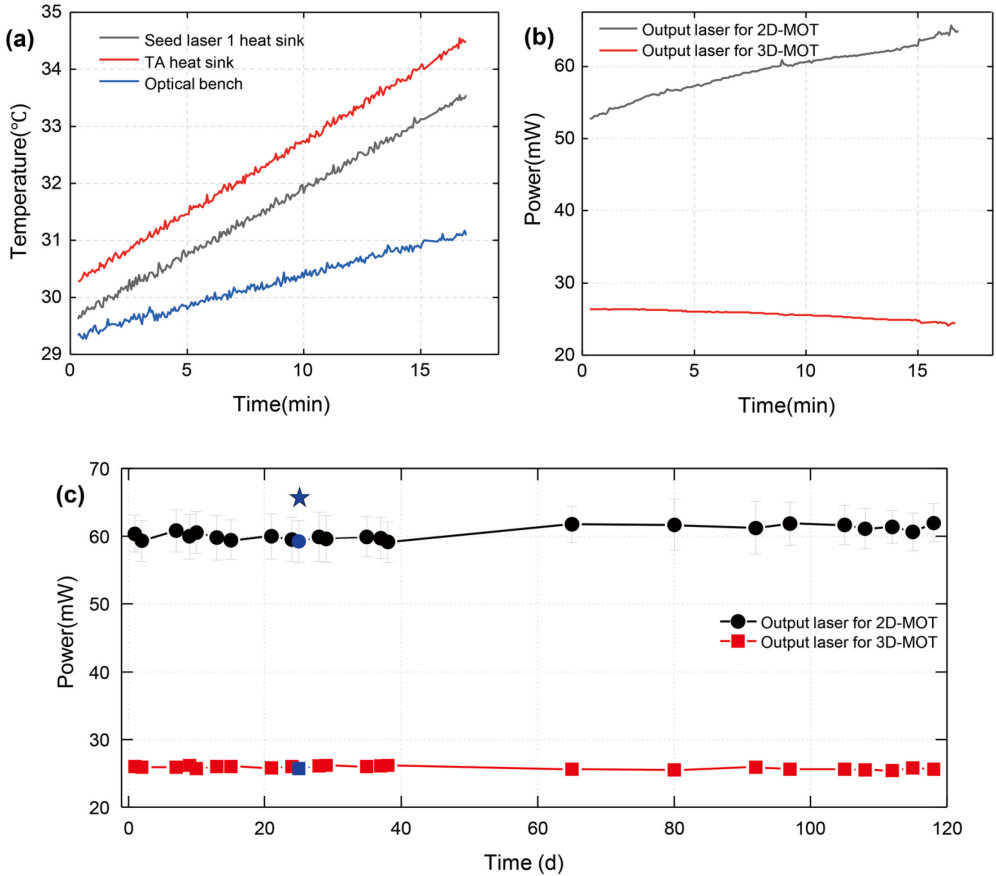

Fig. 10. The power variations of output lasers of the optical system in space. (a) Temperatures recorded by detectors of the optical system during one scientific experiment. (b) Power variations of output the lasers for the 2D-MOT and 3D-MOT during one scientific experiment. (c) Recorded powers of the output lasers for the 2D-MOT and 3D-MOT for 4 months. (a) and (b) correspond to the data labeled with a blue star in (c).

## 6. Conclusion and discussion

The CSSAI is a dual-species AI payload for operation in CSS. Its optical system must be designed to satisfy requirements for complex functions, compact size, good thermal and mechanical stability, and high reliability, which presents significant design challenges. This article introduces the design, analysis, and tests of this optical system. To simplify the scheme design, we propose a method that combines sideband frequency locking, sideband modulation, and synchronous power amplification. Under this design, the system achieves all necessary functionalities for dual-species AI operations using only 2 seed lasers, 1 TA, and 1 AOM. For the mechanical design, the fused silicon is employed as the optical bench, and bond technology is used for its compact optical components. Vibration isolation and thermal conductivity designs are implemented to enhance the thermal and mechanical performance of the optical system. The assembled optical system has a volume of only 6.2 l and a weight of 5.2 kg, which is the most compact one among typical reported optical systems for both ground-based and space-based AIs [34, 38, 39, 42, 44, 46, 48]. The optical system also passed strict thermal and mechanical tests to verify its environmental adaptability. Finite element analysis is applied to calculate the variations of the output lasers caused by the temperature and gravity changes. The functions and parameters of the optical system are tested and measured both on the ground and in space. Space-based laser cooling and cold atom interference are realized, which verifies that the optical system meets its design requirements. The laser powers are recorded for months and have a fluctuation of less than 2.5%, which verifies the long-term stability of the laser system in space.

Further research could be focus on the optimization of the structural design to further enhance the mechanical and thermal stability of the optical system. The design of the optical system introduced in this article could serve as a valuable reference for the design of the compact cold atom payloads in space, such as atom gravimeters, atom gyroscopes, cold atom clocks, and other cold atom instruments.

**Funding.** Space Application System of China Manned Space Program Second batch of the Scientific Experiment Project (04-3-02); Space Application System of China Manned Space Program (JC2-0576); Innovation Program for Quantum Science and Technology (2021ZD0300603, 2021ZD0300604); the Hubei Provincial Science and Technology Major Project (ZDZX2022000001); Defense Industrial Technology Development Program (JCKY2022130C012); National Natural Science Foundation of China (12174403, W2412045, 12204493); Wuhan Dawn Plan Project (20230102010202082); China Postdoctoral Science Foundation (2020M672453).

**Acknowledgment.** We would like to thank the support and advice from Hongen Zhong from Technology and Engineering Center for Space Utilization of CAS, Tang Li and Qiuzhi Qu from Shanghai Institute of Optics and Fine mechanic, and Rong Shu from Innovation Academy for Microsatellites of CAS, and also the supports for the in-orbit experiments from Shan Jin from the Payload Operation and Application Center of the CSS and Zongfeng Li from Project team of High Microgravity Level Research Rack, and many others.

**Disclosures.** The authors declare no conflicts of interest.

**Data availability.** Data underlying the results presented in this paper are not publicly available at this time but may be obtained from the authors upon reasonable request.